\documentclass[journal]{IEEEtran}
\pdfoutput=1
\usepackage{amsmath}
\usepackage{algorithmicx}
\usepackage{algorithm}
\usepackage{algpseudocode}
\usepackage{color}
\usepackage{pxfonts}
\usepackage{xcolor}
\usepackage{graphicx}
\usepackage{caption}
\usepackage[caption=false,font=normalsize,labelfont=sf,textfont=sf]{subfig}
\usepackage{float} 
\usepackage{multirow}
\usepackage{enumerate}
\usepackage{diagbox}
\usepackage{bm}
\usepackage{stfloats}
\usepackage[labelfont=bf]{caption}
\usepackage{booktabs}
\usepackage{hyperref}
\hypersetup {colorlinks,allcolors=black}
\usepackage{comment}
\usepackage{epstopdf}

\makeatletter
\let\c@lofdepth\relax
\let\c@lotdepth\relax
\let\subfloat@label\relax
\let\c@subfigure\relax
\let\l@subfigure\relax
\let\c@subtable\relax
\let\l@subtable\relax

\let\@listsubcaptions\relax
\let\@dottedxxxline \relax
\let\sf@@sub@label\relax

\makeatother
\usepackage{subfigure}

\begin{document}

\title{Autonomous Smart Grid Fault Detection}

\author{Qiyue Li,~\IEEEmembership{Senior Member,~IEEE,}
        Yuxing Deng,
        Xin Liu,
        Wei Sun,~\IEEEmembership{Senior Member,~IEEE,}
        Weitao Li,
        Jie Li,~\IEEEmembership{Member,~IEEE,}
        Zhi Liu,~\IEEEmembership{Senior Member,~IEEE}

\thanks{Qiyue Li, Yuxing Deng, Wei Sun and Weitao Li are with the School of Electrical Engineering and Automation, Hefei University of Technology, Hefei, China, and also with the Engineering Technology Research Center of Industrial Automation of Anhui Province, Hefei, China (e-mail: liqiyue@mail.ustc.edu.cn; yxdeng@mail.hfut.edu.cn; wsun@hfut.edu.cn; wtli@hfut.edu.cn).}
\thanks{Xin Liu is with the State Grid Anhui Ultra High Voltage Company, Hefei, China (e-mail: hfutlx@hotmail.com).}% <-this % stops a space
\thanks{Jie Li is with the School of Computer Science and Information Engineering, Hefei University of Technology, Hefei, China (e-mail: lijie@hfut.edu.cn).}
\thanks{Zhi Liu is with The University of Electro-Communications, Japan (e-mail: liuzhi@uec.ac.jp).}% <-this % stops a space
\thanks{Corresponding author: Zhi Liu.}}

\maketitle

\begin{abstract}

Smart grid plays a crucial  role for the smart society and the upcoming carbon neutral society.
Achieving autonomous smart grid fault detection is critical for smart grid system state awareness, maintenance and  operation. This paper focuses on fault monitoring in smart grid and discusses the inherent technical challenges and solutions. In particular, we first present the basic principles of smart grid fault detection.
Then, we explain the new requirements for autonomous smart grid fault detection, the technical challenges and their possible solutions.
A case study is introduced, as a preliminary study for autonomous smart grid fault detection.
In addition, we highlight relevant directions for future research.
\end{abstract}

\begin{IEEEkeywords}
smart grid, autonomous fault detection, machine learning, resource allocation, cloud computing, edge computing
\end{IEEEkeywords}

\section{Introduction} \label{sec_introduction}

Smart grid is a power network that enables a two-way flow of electricity and data through digital communication technologies that can detect, react and proactively act on changes in usage and multiple issues. %
It plays a crucial role for the smart society and the upcoming carbon neutral society. 
The core technologies in smart grid include the sensing, communication, optimization and data analysis for various purposes \cite{fang2011smart}.

Fault detection of smart grid is an important research problem that has attracted increasing attention from both academia and industry. It is essential to improve the performance and reduce disruptions of smart power systems.
The presence of faults or incipient faults in smart grid can be determined by measuring and analyzing changes in electrical power (e.g., current and voltage), environmental and equipment parameters, a process known as smart grid fault detection. 
Achieving autonomous smart grid fault detection is crucial to system status awareness, maintenance and operation.

With the continuous innovation of power systems and equipments, and the carbon neutral goals, more devices and technologies are used for fault detection. All these make fault detection in autonomous smart grid challenging.
\begin{itemize}
    \item More complex operating conditions of new power system have put forward higher requirements for the state awareness and operation maintenance of power equipments: 
    
    The operational safety of power equipments is the basis for reliable operation of the entire power grid. Comprehensive, timely and accurate sensing and adaptive adjustment is a prerequisite for ensuring the safety of power equipments \cite{9134959}. With a high proportion of renewable energy accommodation, the new power system with strong uncertainty, high volatility and many harmonics will lead to more extreme and variable operating conditions, which puts forward higher requirements for safe and reliable operation \cite{9163315}. It is necessary to solve the problems of real-time sensing, accurate assessment of equipment status and timely warning of potential faults under complex and variable conditions, and to study the strategy of long-term service maintenance, so as to realize lean management and efficient maintenance of power equipment and ensure the safety and reliability of long-term operation of power equipment under the complex operating conditions of new power system.

\item The massive adoption of new power equipments has brought new challenges to operation and maintenance:

New power equipment are the key infrastructures used to support the construction and development of new power systems, such as power conversion and control devices based on power electronics technology, large-scale energy storage equipment, offshore wind power access-related equipment, etc. These new types of power equipment have a relatively short history and are used in relatively few traditional power grids \cite{9436032}. They lack effective condition assessment methods that have been tested in practice \cite{8846092}. The operation and maintenance technology of the equipment is still immature, and basic scientific issues such as the mechanism and evolution law of equipment failure need to be studied in depth.

\item Carbon neutrality puts forward new requirements for the operation of power equipments:

Green and high efficiency is one of the main features of the new power system \cite{8804208}. While the power grid strongly supports the access and consumption of renewable energy resources, it should further reduce the carbon emission level of the power grid itself, and realize the low-carbon transformation of planning, operation and maintenance. Considering the huge number and rapid growth of power equipments, improving the utilization efficiency of existing power equipment and reducing equipment operating losses are important goals for the efficient operation of new power systems. The key issues that need to be studied mainly include real-time identification and accurate prediction methods of key parameters that affect the utilization and service life of equipment, factors affecting the comprehensive energy consumption of power grid equipment, and control strategies.
\end{itemize}

%Smart  grid  fault  detection  is  used  to  determine whether  there  is  any  cause  to  the  smart  grid  fault  by measuring and analyzing the electricity (such as current and voltage) and the change in the switching value  of  the  protection  circuit  breaker  operation \cite{8004464}. 

%\textcolor{red}{delete this?}？In the smart grid, measuring electricity requires the use of Smart Sensors(SSs), which are sensing devices with digitalization capacity and digital information processing functionalities. SSs may communicate using standardized communication protocols, such as the IEEE 1451 family of Smart Transducer Interface Standards, IEEE 1815 Standard for Electric Power Systems Communications - DNP3, IEEE C37.238 PTP Power Profile, and others.Analysis of SS data is an important part of smart grid fault detection.

%\textcolor{red}{some discussions on the state-of-the-art}

To the best of our knowledge, there are still many unsolved research issues in autonomous smart grid fault detection.
To this end, this paper first presents the basic principles of smart grid fault detection. Then, we explain the new requirements for autonomous smart grid fault detection, the technical challenges and their possible solutions. A case study is conducted and shows the great advantages of the proposed solution. Furthermore, we highlight relevant directions for future research.

This paper is structured as follows. Section \ref{smart_grid_and_smart_grid_fault_detection} introduces smart grid and smart grid fault detection. Section \ref{Challenges_and_Solutions} introduces the challenges and solutions for autonomous smart grid fault detection. One actual case deployed in the ultra high voltage (UHV) converter substation with the highest voltage levels in the world is presented in Section \ref{case_study}. In Section \ref{future_work}, we look ahead to the problems that need to be solved in the future. Finally, the manuscript is concluded in section \ref{sec_conclusion}.

\begin{figure*}
    \centering
    \includegraphics[width=7in]{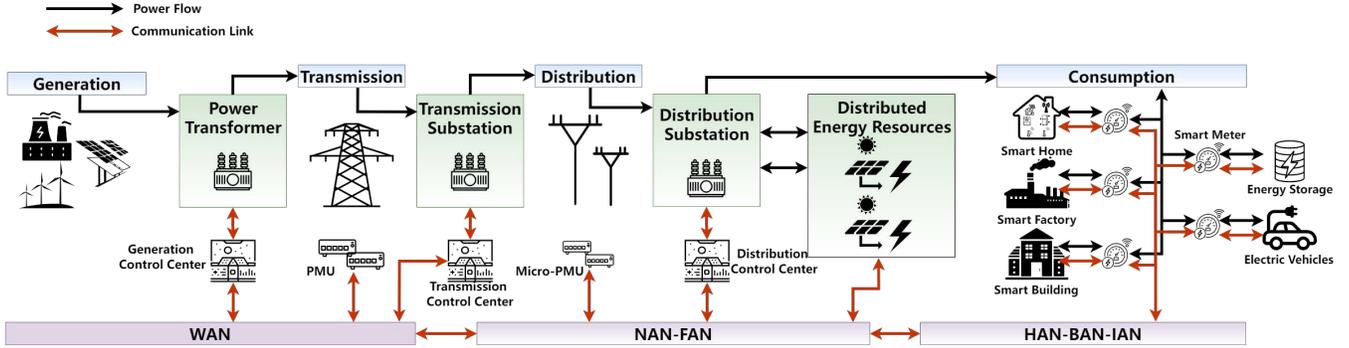}
    \caption{Key components of smart grid system. From power generation, power transmission, power distribution including distributed energy resources, to power consumption, the status of all the hardware infrastructures needs to be sensed, transmitted by different networks, processed by local or cloud control center, to evaluate the safety and detect the fault of power grid.}
    \label{fig_smart_grid}
\end{figure*}

\section{Smart grid and smart grid fault detection}
\label{smart_grid_and_smart_grid_fault_detection}
In this section, we overview the key components of the smart grid and smart grid fault detection.

\subsection{Overview of Smart Grid and Fault Detection}

The key components of smart grid system is shown in Fig.\ref{fig_smart_grid}. From the perspectives of power transmission, power distribution and power consumption, autonomous smart grid fault detection is needed.

\subsubsection{Power Transmission}

As UHV AC and DC transmission systems become larger, more complex and more intelligent, the operating characteristics of power grid have undergone profound changes. On one hand, UHV power transmission is affected by factors such as changing regional climate and complex operating environment, and it is urgent to control the operation situation of the power grid synchronously. On the other hand, the transmission and reception of UHV power grids are closely coupled with AC and DC, which can easily lead to global security risks. Therefore, real-time early cross-regional  risk warning and fault detection at the network-wide level is extremely important to assist in scheduling operations.

% 随着中国特高压交直流输电系统日趋大型化、复杂化和智能化，电网运行特性随之发生深刻变化，电网调度也面临着新的重大挑战。一方面，特高压输电受到地区气候多变、运行环境复杂等因素的影响，迫切需要同步掌控电网运行态势；另一方面，特高压电网送受端、交直流耦合密切，容易引发全局安全风险。因此，在全网层面实时进行跨区域安全预警和风险防范控制对协助调度运行极为重要。
% 特别是不同于传统交流电网的交直流混连输电系统发生输电线路故障时，由于换流器的非线性时变特性以及直流控制系统的复杂性，换流站两侧交、直流系统的电气量存在强耦合关系。交流线路故障时，会导致直流侧电气量发生变化，从而改变直流侧的故障特征，进而影响直流保护性能。故障严重时甚至会引发直流系统逆变器换相失败，直流控制系统的调节过程会对交流系统产生二次冲击，容易引起交流保护误动作。因此，在交直流输电系统中，仅依据保护动作信息无法及时准确地诊断出真实故障。现阶段交直流输电系统的故障诊断过程中需要高效获取准确、完备的诊断信息。

\subsubsection{Power Distribution}
With the increase of penetration rate, a large number of distributed generations, electric vehicle charging devices, energy storage devices, and microgrids are connected to the distribution network. The randomness of renewable energy sources and loads, as well as the distributed multi-agent control characteristics, are likely to cause large power fluctuations and voltage over-limit, which make the operation of the distribution network face severe challenges. Meanwhile, the tie switch based network reconfiguration method is limited by the response speed, operational life and inrush current, cannot meet the needs of future high-reliability users.
%大量的分布式能源(distributed generation，DG)、电动汽车充电装置、储能装置以及微网接入配电网，而且渗透率不断提高。新能源和负荷的随机性以及多主体利益特征下分散控制特性，容易引起馈线功率大幅度波动以及电压越限，使配电网的运行面临严峻挑战。同时，基于联络开关的网络重构受到开关响应速度、动作寿命和冲击电流等问题的限制，不能满足未来高可靠性用户的需求。  DS (Digital signal)？

%In particular, the presence of multiple fault current contributing sources, changing fault levels, bi-directional power flows, and formation of sub-grids within the DS for improved reliability. Due to these reasons, there is a possibility of coordination failure of protective relays, which can lead to wrong relays being operated. Hence, traditional impedance-based fault location schemes and fault location schemes based on monitoring the operating status of protection devices are becoming obsolete [1]. Further, it is not possible to completely rely on substation measurements in future DS as emphasis is being made to operate DS as a group of smaller islands for technical and economic reasons. In such circumstances, it is preferable to use the data from monitoring devices (like smart meters, IEDs, PMUs) distributed across the DS than from a single location like substation.

\subsubsection{Power Consumption}
%    随着用电信息采集的深度推进，为供电电压监测、电能质量管理、用电稽查提供了专业的分析数据。用电信息采集稳定运行、采集信息量完整可靠显得尤为重要，加强运行维护工作是保障采集稳定运行及各项考核指标提升的必然途径.
%    由于用电信息采集建设范围广、工作量大，系统涵盖采集主站、上下行通信信道(230 MHz, GPRS,CDMA、光纤、载波、RS485) 、采集终端、智能电能表各业务处理、通信、计量、远程费控等诸多环节。在用电信息采集快速稳步推进过程中，对实施、运维的工作量、工作质量及效率提出了更高要求，能够实现自主的对用电采集数据进行分析检测十分重要。
The structure of low voltage consumption power grid is extremely complex. Electrical faults, such as leakage fault, short circuit fault, overload fault and large contact resistance, are prone to occur during operation, which affect the users' daily applications. With the in-depth advancement of power consumption information collection, it provides professional analysis data for power supply voltage monitoring, power quality management, and power fault detection. Strengthening the operation and maintenance work is an inevitable way to ensure the stable operation of the collection system and the improvement of various assessment indicators.

%Due to the wide range of power consumption information collection and the large workload, the collection system covers the collection master station, uplink and downlink communication channels (230 MHz, GPRS, CDMA, optical fiber, carrier, RS485), collection terminals, smart meters and many other links. In the process of rapid and steady progress in the collection of power consumption information, higher requirements are put forward for the workload, work quality and efficiency of implementation, operation and maintenance, and it is very important to be able to automatically analyze and detect power consumption data.

\subsection{Advanced Technologies in Autonomous Fault Detection}
\begin{figure*}[!htb]
    \centering
    \includegraphics[width=5in]{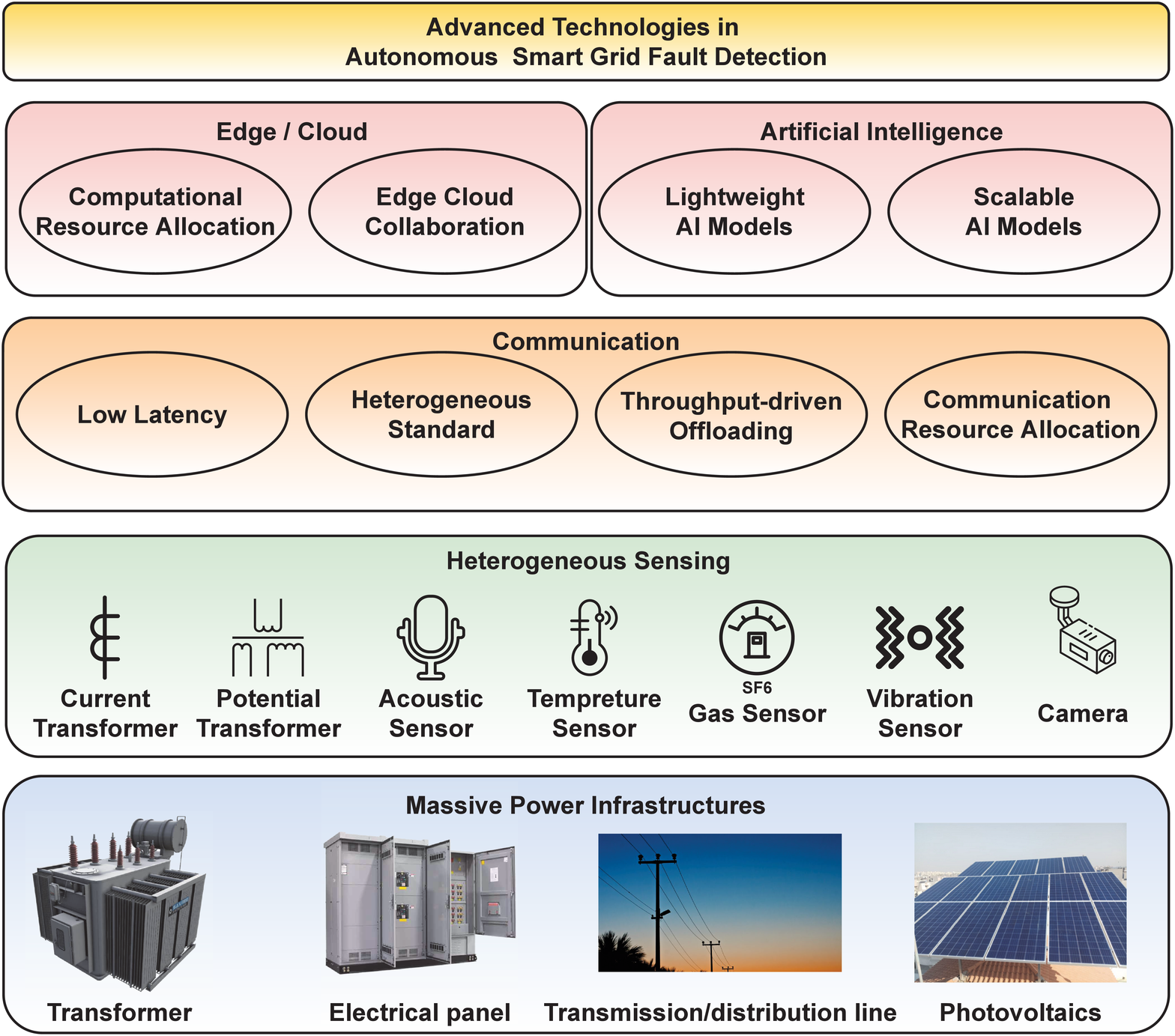}
    \caption{Advanced technologies in autonomous smart grid fault detection.}
    \label{fig_advanced_technology}
\end{figure*}

As illustrated in Fig. \ref{fig_advanced_technology}, we describe some advanced technologies playing important roles in autonomous smart grid fault detection.

\subsubsection{Advanced Sensing and Communication}
%一段话介绍Smart Grid Fault Detection中需要的新技术，包括检测（除了传统电流和电压互感器之外的新型检测方式）、通信与传输（多种传输方式，本地与云端，突出一下5G、6G中的URLLC、mMTC等新技术的应用可能）、处理（AI的方法进行Fault Detection，包括edge intelligence和cloud intelligence）

With the continuous development of smart grid, its fault detection technology is also constantly developed and updated. In terms of sensing technology, in addition to traditional current and voltage transformers, sensing devices such as phasor measurement units, smart meters, acoustic sensors, vibration sensors, environmental sensors, visible and infrared cameras have also been deployed in the grid system. Massive data collected by sensors needs to be sent to cloud servers or edge devices for processing. At this time, low latency communication with cloud servers and edge devices also needs to be considered. In the smart grid, there are many communication standards (e.g., Ethernet, optical fiber Ethernet, LTE, WiFi, ZigBee). With the realization of 5G technology and throughput optimized communication resource allocation, Ultra-Reliable Low-Latency Communications (URLLC)  and massive Machine-Type Communication (mMTC) makes data transmission more reliable and faster, and provide support for real-time fault detection. 

%it also includes a large number of semi-structured and unstructured data, such as text recording, audio data, partial discharge spectrum, infrared image, line patrol image, monitoring video, frequency response and sampling waveform. 
%Moreover, with the development of artificial intelligent (AI) and the emergence of lightweight neural networks, neural networks can run in the cloud and edge devices for data processing, realizing cloud intelligence and edge intelligence.

\subsubsection{Artificial Intelligence and Machine Learning}
%The artificial neural network is composed of the interconnection of neurons (perceptrons). There is a functional relationship between the input and output of the neuron, and this function is called the excitation function. The excitation function produces an output only when the information input to the neuron exceeds a certain criterion. A multilayer perceptron is the simplest neural network, including an input layer, a hidden layer (intermediate layer), and an output layer. Each layer is composed of multiple neurons, and the neurons are fully connected. The neural network model is trained through a large amount of sample data, so that the data input of the network in the real environment can get an output, so that the loss function (cost function) is minimized.
%人工神经网络由神经元（感知器）的互联组成网络，神经元的输入和输出之间具有函数关系，这个函数称为激励函数。只有当输入神经元的信息超过一定标准时，激励函数才会产生输出。多层感知机是一种最简单的神经网络，包含输入层、隐含层（中间层）、输出层，每一层由多个神经元构成，且神经元之间具有全连接的特性。通过大量样本数据对神经网络模型进行训练，使网络在真实环境下的数据输入都可以得到一个输出，使得损失函数（代价函数）最小。

%Powerful Artificial intelligence algorithms can deeply integrate and analyze massive online and offline multi-source heterogeneous data of power systems. And extract the mapping relationship, key information and judgment rules to establish a data-driven prediction, evaluation and diagnosis model, which can improve the real-time and accuracy of system fault diagnosis and prediction.
Now advanced artificial intelligence (AI) technology has entered the stage of independent learning relying on big data, and with further development, it does not require manual control throughout the process. This is particularly evident in the work of power system fault detection. 

The operation structure of conventional power system is complex, with extremely cumbersome internal links. If the power system fails during operation, it will be very difficult and error-prone to use the traditional manual troubleshooting methods. With the popularization and improvement of AI technology, machine learning algorithms can be used to deeply integrate and analyze the massive online and offline multi-source heterogeneous data of power systems, and extract the mapping relationships, which can quickly, efficiently and accurately detect and intelligently handle most power system faults. Different models are suitable for processing different types of data, and deep learning can also integrate multi-source heterogeneous data, which greatly improves 
the efficiency and ensures safe and reliable operation of the power grid.

\subsubsection{Edge Computing and Cloud-Edge Collaboration}
Despite the advances in AI and computing technologies, there lies the need for instant availability of the required computational resources due to the ever-increasing number of smart devices and the evolution of complex machine learning algorithms. With the help of communication and cloud computing, massive data processing and storage can be performed in cloud servers. However, cloud computing-based systems also 
havelimitation of network latency with high dynamic jitter, which in turn hinders the stringent delay requirement of fault detection in smart grid.

Edge computing has recently emerged as a way to reduce the computing pressure of the end devices and the total delay of transmission and monitoring. 
By shifting part of the work in cloud servers to edge devices, a cloud-edge collaboration system can be formed. Through the study of data scheduling and computational resource allocation, the performance of such system can meet the delay bound requirements of fault detection and process massive data at the same time.

\section{Challenges and Solutions in Autonomous Smart Grid Fault Detection}
\label{Challenges_and_Solutions}

This section overviews the technical challenges and possible solutions for autonomous smart grid fault detection.

\subsection{Heterogeneous Sensing}

%更复杂的智能电网、更高的可靠性要求，需要多种检测方式（时序数据与图像数据）
Sensing data is the basis for evaluating the status of power grid hardware infrastructures. According to the sampling frequency, it can be divided into three categories: static data, dynamic data and quasi-dynamic data. Static data mainly includes equipment account, nameplate parameters, test data before commissioning, geographic location, etc. Dynamic data is usually updated periodically by seconds, minutes or hours, mainly including equipment online monitoring data, operation data, inspection records, live detection data, environmental meteorological data, etc. And the quasi-dynamic data is usually updated regularly or irregularly on a monthly or yearly basis, mainly including fault defect data, equipment hidden danger records, and maintenance records.

With the accumulation of a large amount of data collected by the power system, the main characteristics of sensing data are as follows:

\begin{itemize}
    \item The amount of data is large and the growth rate is fast. The types of electrical equipment is varietal, and the number is huge. Besides with the deployment of a large number of online monitoring sensors, the growth trend of condition monitoring data volume is becoming more and more obvious.
    \item The sources of sensing data are heterogeneous. With the continuous development of digital power grids, information platforms such as power equipment status monitoring systems, energy management systems, geographic information systems, and meteorological systems have been widely used, making the relevant information sources of sensing data more extensive.
\end{itemize}    

\textbf{Solution:} To solve the challenges of heterogeneous sensing, one possible solution is data fusion. A large number of multi-source sensors will collect various information of power equipments, some of which are related to each other, while some parameters are not affected by others. Data fusion methods can automatically analyze and optimize information according to certain criteria, to obtain a consistent and complete characteristics of the equipments, and complete the required logical processing and decision-making. Through information-level fusion, feature-level fusion and decision-level fusion, the redundancy of the heterogeneous sensing data will be eliminated, which can improve the accuracy of fault detection.

%and the information will be \textcolor{red}{hhhh 后面被截断了。}It is effectively processed by fusion,  

%The heterogeneity of sensing data is mainly manifested in the diversification of data structures. In addition to structured data, These heterogeneous sensing data contains a large number of real-time operation information and maintenance history data of power equipment, from which fault status characteristics such as partial overheating, partial discharge, and mechanical property damage of power equipment can be mined.

%电力设备数据是评估设备状态的基础,根据电力设备状态信息的更新频率,可将设备状态数据分为3类：静态数据、动态数据和准动态数据。静态数据主要包括设备台帐、铭牌参数、投运前试验数据、地理位置等;动态数据通常按分钟、小时或天为周期更新,主要包括设备在线监测数据、运行数据、巡检记录、带电检测数据、环境气象数据等;准动态数据通常按月或年定期或不定期更新,主要包括故障缺陷数据、检修试验数据、设备隐患记录、检修记录等。电力设备状态数据的来源如表2所示。
%随着电力系统信息化平台采集数据的大量累积,电力设备运维状态数据所呈现的主要特征如下[9,43-44]：
%1）数据体量大,增长速度快。电力设备类型多,数量庞大。另一方面,随着在线监测传感器的大量部署,状态监测数据体量的增长趋势愈发明显。
%2）设备数据来源广泛。随着数字化电网的不断发展,电力设备状态监测系统、能量管理系统、地理信息系统及气象信息系统等电力信息化平台逐步完善并广泛应用,使得电力设备运行状态数据的相关信息来源更加广泛,数据模型、格式和接口也更加多样。
%3）数据类型异构多样。设备数据的异构性主要表现在数据结构的多样化上,除去结构化数据外,还包括如文本记录、音频数据、局部放电图谱、红外图像、巡线图像、监控视频、频响和波形曲线等在内的大量半结构化和非结构化数据。这些异构的设备状态数据中包含了大量电力设备实时运行信息和检修维护历史数据,从中可以挖掘电力设备局部过热、局部放电、机械特性受损等故障状态特征。

\subsection{Stringent Delay Requirement}
%\subsection{Data Scheduling and resource allocation in Cloud-Edge Collaboration}
%\item 2. 云边协同的方式，需要数据调度、通信资源分配、计算资源分配
%Edge computing \cite{shi2016edge,guo2017energy} has recently emerged as a way to reduce the computing pressure of the cloud and the total delay of communication and monitoring. 
%By shifting part of the work in clouds to edge devices, a cloud-edge collaboration system can be formed.
%Through the study of data scheduling and resource allocation in this type of system, communication delay and resource consumption has been reduced. \cite{8664595} aims at minimizing the weighted-sum delay of the devices, and studies the communication and computational resource allocation problems in the cloud-edge collaborative system. \cite{8491367} developed a method that considers content caching and computational resource allocation at the same time, which can minimize the total delay of mobile edge devices.
Autonomous smart grid fault detection has stringent delay requirement.
However, the current studies of cloud-edge collaborative smart grid detection do not fully consider the delay requirement and the flexibility of the detection, and the resource allocation is complicated.
%
%As a result of these advantages, edge computing-assisted smart grid monitoring and detection have been studied, particularly to reduce the delay \cite{8640019}, \cite{7588230}, \cite{8849012}. Note that the hard delay limits required by the smart grid are not considered in these schemes, and thus, the detection performance cannot be guaranteed. Moreover, they do not provide flexible detection services, which degrade the system performance, due to detection demands and the need to optimize the available resources. 
Edge computing can provide relatively low latency. However, the limited computational resources may not be able to take up the task of local detection of all data and the computation time cannot be neglected.
%Edge computing can provide relatively low delay. However, the limited computational resources and may be unable to bear the local detection task for all data, and the computing time unable to be ignored,
%Moreover, considering that most edge devices have limited computational resources and may be unable to bear the local detection task for all data, and the computing time unable to be ignored, 
Thus, lightweight neural networks that consume different computational resources, computing times, and provide different detection accuracy can be designed for fault detection.
The inherent selection of neural network with different detection accuracy and complexity makes the resource allocation even more complicated.
To the best of our knowledge, these issues remain unsolved.

\textbf{Solution:} For smart grid fault detection with delay and accuracy requirements, one possible solution is a cloud-edge collaborative fault detection system, which runs a high-performance neural network in the cloud serve and a low-precision lightweight neural network on the edge device to achieve fault detection.
The inherent research problem is how to solve the resource allocation problem of such systems in an efficient and timely manner, either by optimization methods or by reinforcement learning-based solutions.

%And a data scheduling and resource allocation algorithm is designed, which can obtain the maximum system throughput while ensuring the delay and accuracy. Solving the algorithm is equivalent to solving a nonlinear integer programming problem, and the optimal solution can be obtained by using the KKT algorithm and the branch-and-bound method.
%the combination of edge computing and intelligence, i.e. edge intelligence, can allow for data to be processed and analyzed locally and intelligently, which effectively reduces the delay and saves bandwidth resources \cite{xu2020edge}. The technical challenge is the most edge devices have limited computational resources and may be unable to bear the local detection task for all data, and the computing time unable to be ignored. Given that there are strict delay limits in the fault detection systems of smart grid, including communication and computing delays \cite{6823238}, it is important to schedule data to get a large throughput while taking into account the stringent delay requirements.
%Thus, lightweight neural networks that consume different computational resources, computing times, and provide different detection accuracy, should be designed for fault detection, and resource allocation and offloading optimization should be studied to improve the performance of the smart grid detection. 
%To the best of our knowledge, these issues remain unsolved.

\subsection{Incipient Fault Detection}

%\item 3. 输电网、配电网和用电网接入的设备种类更多，高效的fault detection更加困难。特别是在配电网与用电网中大规模分布式电源的接入，改变了传统故障识别使用的信号特征，更难以识别的incipient fault需要high performance AI neural network赋能。
%\textcolor{red}{liu:下面这几个关于fault类型的讨论是否可以放到上面一个section??}
%\textcolor{red}{liu:这样这段大概说说检测有啥难的，比如持续时间很短等等，加上资源有限，是否可以把lightweight合在这个地方}

Before the power equipment fails, some pre-occurring abnormalities called incipient faults will occur. %\cite{7587342}
It is usually a self-cleaning fault, which lasts for a short time and recurs. %\cite{9432388}.
%In the distribution network, the causes of these faults are different, and the locations of occurrences are random. 
%Incipient fault detection allows maintenance personnel to replace defective equipment in advance, effectively improving power supply reliability. In addition, it is a predictive maintenance, and detecting failures in the early stages helps avoid unexpected interruptions and subsequent losses \cite{9432388}.
Incipient faults are usually accompanied by uncertain and random arcs, and the duration ranges from a quarter cycle to up to four cycles \cite{5424106}, making incipient faults difficult to analyze.
Most of the traditional incipient fault detection methods analyze the characteristics of the waveform and then classify the fault. %\cite{5382504} proposes a numerical modeling method to establish incipient failure modes from oscillometric data, and uses self-organizing map technology to extract features in the wavelet domain for fault detection. 
%\cite{7576142} proposed a method to detect multi-cycle incipient faults, which is performed by calculating the total harmonic distortion of the fault point voltage at all possible distances in the cable. 
For example, a human-level concept learning based method is illustrated to detect incipient failures in \cite{article}, in which two steps, human-level waveform decomposition and hierarchical probability learning, are involved.
However, these manually selected features have insufficient representation ability and classifiability for incipient faults, leading to poor fault detection performance. In addition, the fault signal in power system presents strong non-stationary characteristics, which makes it difficult to set appropriate metric of autonomous smart grid fault detection.

%At this time, AI plays a role.

Using AI methods can quickly detect incipient faults and easily respond to the changes in waveform characteristics. For example, \cite{8744581} proposes a hybrid method of initial fault monitoring and classification, which uses machine learning algorithms to detect the faults. However, incipient faults rarely occur, and its duration is very short. When they appear, the protective devices deployed in power system may not operate, and the fault recorder will not be able to record the signals. Therefore, not enough training samples can be obtained to train an accurate AI models.

\textbf{Solution:} One possible solution to the incipient fault detection is to extract time and frequency characteristics of the non-stationary signal, which can be achieved through wavelet transform. Then the time-frequency information can be embedded into recurrent neural networks to achieve better performance in non-stationary signal analysis. Besides, data augmentation methods, such as faulty signal switch, temporal extension, generative adversarial network (GAN), semi-supervised learning, can be designed to solve the problem of few training samples.

\subsection{Resource Constrained Edge Computing}
%\item 4. 边缘智能的轻量化
%Traditional routine inspection and maintenance of transmission lines are performed manually. Due to heavy workload, high degree of difficulty, low efficiency, and poor precision, it is difficult to meet the actual requirements of new power systems. It is, therefore, necessary to develop efficient inspection methods.
Edge computing can reduce the computing pressure of the cloud servers and the total delay of communication and monitoring \cite{shi2016edge}. The combination of edge computing and intelligence, i.e., edge intelligence, can allow the sensing data of power system to be processed and analyzed locally and intelligently, which effectively reduces the delay and saves bandwidth resources. However, most edge devices have limited computational resources, and may not be able to take on the task of local detection of all data with negligible computing time. The accurate but large AI models for fault detection with billion parameters cannot deployed on the resource constrained edge devices. 

Furthermore, given that there are strict delay constraints in the  smart grid fault detection systems, including communication and computing delays, it is important to schedule data to get a large throughput while taking into account the stringent delay requirements \cite{6823238,li2021optimal}. However, the data scheduling and computational resource allocation can be formulated into a mixed integer optimization problem, which is NP-hard. Solving such problem usually takes a long time and cannot meet the delay requirements either.

\textbf{Solution:} One possible solution of the resource constrained edge computing is scalable AI models. To meet the delay requirements, we can design several lightweight neural networks with different structures, computing time, consumed resources, providing different detection accuracy. By allocating appropriate computational resources of edge device, the system throughput can be maximized and resource utilization can be improved while meeting the requirements of data transmission and processing delays. Besides, we can utilize deep reinforcement learning methods to solve the NP-hard data offloading and computational resource allocation problem in quasi-real-time with near optimal performance.

%Transmission line fault detection based on artificial intelligence is mostly realized by neural network. Generally speaking, the more complex the neural network model is, the higher the detection accuracy is, and the more computational resources are consumed. To obtain a lightweight neural network that can run on embedded devices, the model can be quantized and pruned to compress the model without sacrificing accuracy, and can also save battery power \cite{gholami2021survey}.
%\textcolor{red}{liu:这个地方可以加个reference来提供几个轻量化的方法}

\subsection{Autonomous Inspection and Fault Detection}
%这里放一个换流站的机器人图

%5. 与电网实际场景的结合（输电系统中可以考虑无人机的边缘侧AI辅助，变电站与换流站中可以使用移动巡检机器人的边缘侧AI辅助，配电、用电中可以使用台区变压器的DTU和用户智能电表的边缘侧AI辅助）

With the development of the power system, the requirements for the safety and reliability of power supply are getting higher. To ensure the normal operation of the power system, power inspection has become one of the important daily tasks of electric utilities. According to the traditional manual methods, the operation and maintenance personnel need to complete the inspection of various equipments in sequence according to the planned route, which consumes a lot of manpower and time.

In the inspection and fault detection of power system, outdoor substations and indoor power equipments are the focus. Regular inspections are required for checking meter data, switch positions, and equipment temperatures at different locations. In addition, the substations and underground pipe corridors have problems such as long inspection distances, strong tunnel closure, inconvenient communication, harmful gases, which pose a certain threat to the personal safety of inspection personnel. Based on this, it has become an inevitable trend from traditional manual inspection to autonomous robots aided inspection and fault detection.

%The traditional electric power inspection adopts the manual method, which not only has low inspection efficiency and high labor cost, but also has missed inspections. In addition, extreme weather such as typhoons, earthquakes, and heavy rains usually pose a great threat to the power grid infrastructure, and manual inspections cannot be carried out due to life safety considerations. In contrast, the environmental adaptability of inspection robots is much higher than that of manual inspections, and robot inspections can be carried out normally in extreme weather.

\textbf{Solution:} 
With the development of information and communication technologies, robots and unmanned aerial vehicles (UAVs) are becoming possible solutions for autonomous inspections and fault detection. In particular, power inspection robots are mainly used in indoor substations, outdoor substations, underground power pipe corridors and other scenarios, while the monitoring of high-altitude power racks, transmission lines, and towers is mainly monitored by drones. On this basis, a complete closed loop of real-time maintenance monitoring of power system can be formed. The mobile robots and UAVs can implement data collection and transmission. After receiving the data, such platforms can also perform fault detection while integrating edge intelligence. The environmental adaptability of the inspection robots is much higher than that of human, saving manpower and greatly improving the quality of inspection.

%compares and analyzes the inspection data, finds equipment hidden dangers and failure symptoms, ensures the stable operation of the equipment, and effectively improves the quality of inspection work. In scenarios such as high-altitude power racks, transmission lines, and tower monitoring, drones are required for monitoring. Through the use of high-precision differential GPS positioning technology, the UAV patrol route can be automatically calculated and planned, and the automatic flight inspection can be automatically eliminated.
%随着电力系统规模的发展，对电力线路的安全运行和供电可靠性的要求越来越高。为了保障电力系统的正常运转，电力巡检已经成为电力从业人员重要的日常工作之一。按照常规传统的人工巡检方式，运维人员都需要按照规划的巡视路线图按照顺序完成各项设备的巡视工作，此项工作耗费大量的人力与时间。

\begin{figure*}[!htb]
    \centering
    \includegraphics[width=7in]{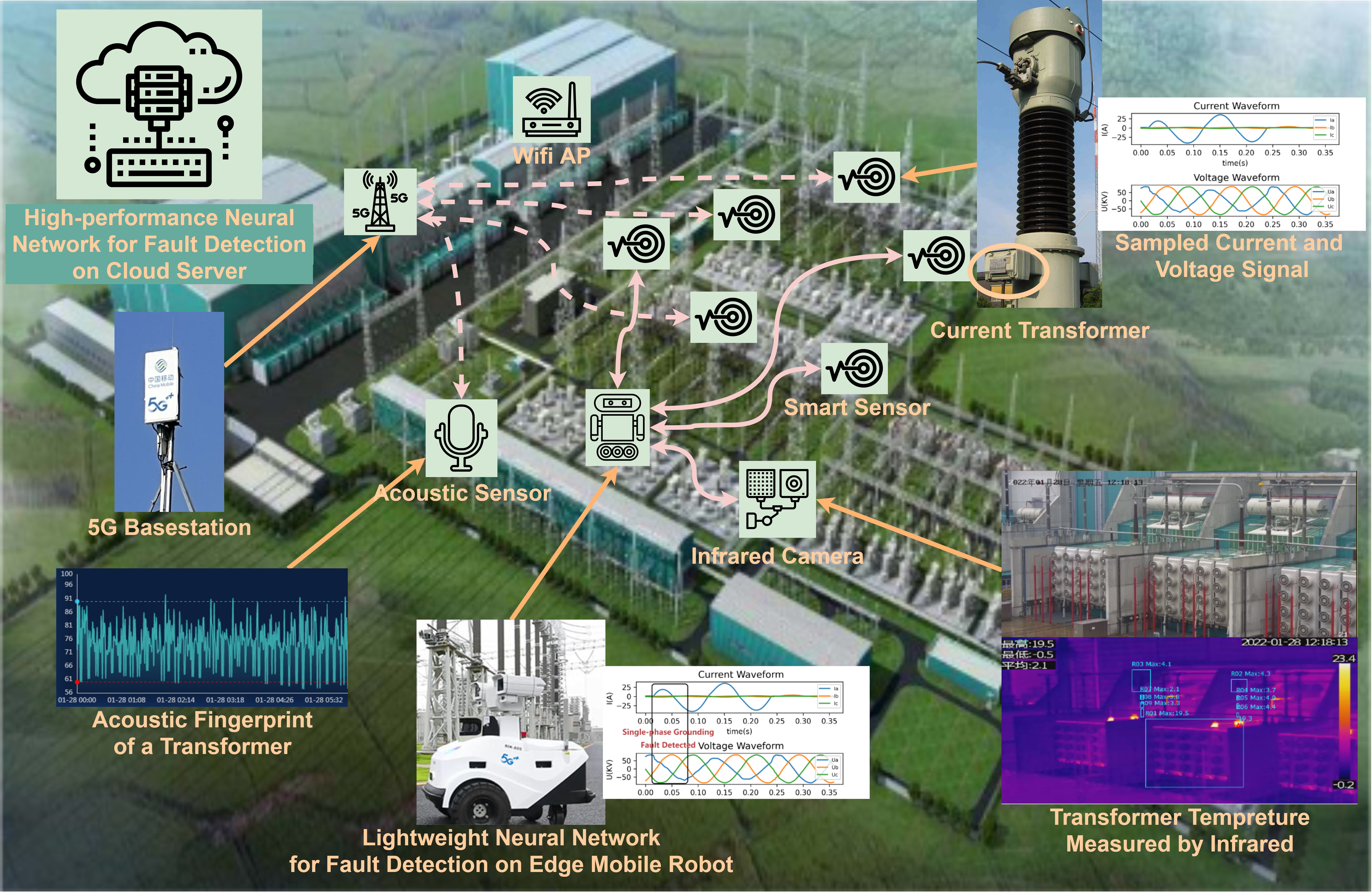}
    \caption{An actual case deployed in the UHV converter substation with the highest voltage levels in the world, Guquan, China.}
    \label{fig_case_study}
\end{figure*}

%在电力巡检中，室外变电站与室内配电设备是监测重点。在变电站等室外场景，针对不同高度和位置的表计读数、开关位置、设备温度等状态均需进行定期检查，而变电站、地下管廊都具有巡检路程长、隧道封闭性强、通讯不便、有害气体存在、环境恶劣等问题，对巡检人员的人身安全会存在一定的威胁。在配电站设备日常人工巡检过程中，关于表计读数、开关位置、设备温度、柜体局部放电等状态检测的工作繁重枯燥。
%基于此，传统人工巡检向机器自动巡检已成必然趋势。传统电力巡检，采用人工方式，不仅巡检效率低、巡检人力成本高，还存在漏检情况；台风、地震、暴雨等极端天气，通常会给电网基础设施带来极大威胁，人工巡检会出于生命安全考虑而无法进行。与之相反，巡检机器人的环境适应性远远高于人工巡检，极端天气下机器人巡检工作也可以正常展开。
%电力巡检机器人主要运用在室内配电站、室外变电站、地下电力管廊等场景，而高空电力架、输电线路、塔架监测主要由无人机监测。在此基础上，形成了一个完整的电力系统实时运维监控闭环。
%室内巡检机器人一般应用在在配电机房、室内变电站等场景。巡检机器人主要实施数据采集和传输，替代人工对室内柜体等设备进行状态检测，接收数据后，平台对巡检数据进行对比和分析，发现设备隐患和故障征兆，保障设备稳定运行，有效提升巡检工作质量。
%在高空电力架、输电线路、塔架监测等场景，则需要无人机配合监测。通过采用高精度差分GPS定位技术，实现无人机巡视航线自动计算规划、自动飞行巡检、自主排除隐患，输电线路交叉跨越测量精度达到厘米级。

\section{Case Study}
\label{case_study}
%\textcolor{red}{liu: case study我觉得3个方面可以加强些。 1。 图4或者说系统介绍的可以更细些，比如为啥她的工作流程，怎么实现了autonoumous。 2. 解法部分要提的多些，比如是复杂的XXX问题，一个思路是XXX解决，另外Branch band啥的也可以更一步帮助。还可以把之前的论文放这，说这个论文又些类似的方法。 3. 结果（目前还没些）要讨论的充分些，比如可以从图里面看到什么insights}

This section introduces the case study, as a preliminary study for the autonomous smart grid fault detection.

\subsection{System Overview}
%In the substation, the distribution network system is complex, and it is very important to detect the fault of the transmission line and power equipment. To ensure the normal operation of the substation, many sensors are deployed in the whole station, which can collect various power data in real time and conduct fault analysis and detection. 
We design a cloud-edge collaborative fault detection system, which is deployed in the UHV converter substation at Guquan, China, which is the receiving end of Changji-Guquan ±1,100 kV UHV DC Transmission Project with the highest voltage levels in the world. As shown in Fig~\ref{fig_case_study}, the system is mainly composed of smart sensors, acoustic sensors, infrared cameras, a 5G base station, an edge intelligence empowered mobile robot, and a cloud server. Among them, the function of the smart sensors is to collect heterogeneous sensing data in real time, such as voltage, current, acoustic fingerprint and vibration signal of transformers, visible and infrared image, as well as other environmental parameters. All the data is transmitted to the cloud server through the 5G base station, or to the edge intelligence robot through WiFi. With a cloud-edge collaboration manner, the high-precision neural network deployed on cloud server and the lightweight neural networks deployed on mobile robot can analyze the sensing data to detect whether there is any faults and the type of faults of the power infrastructures in the substation.

%The high-performance LSTM neural network runs on the cloud server, and the lightweight LSTM neural network runs on the edge mobile robot, realizing cloud intelligence and edge intelligence. The sensors collect data in real time and send it to the cloud server or edge intelligence empowered mobile robot, and the neural network in the cloud server or edge intelligence empowered mobile robot analyzes the data in real time, realizing autonomous fault detection.
Although 5G network with low air interface latency is used, the delay of transmitting data to cloud server through backbone network remains significant and is highly dynamic and jittery, which may exceed the stringent delay bound requirement. 
Moreover, the computational resources of mobile robots are limited, and the computation time is not negligible even when a lightweight neural network is deployed. To this end, we design an algorithm for data scheduling and computational resource allocation in the cloud-edge collaborative system, which can obtain the maximum system throughput while ensuring accuracy and delay, taking real-time network delay, available computational resource at edge servers and detection accuracy of neural networks into account. The process of scheduling and resource allocation is equivalent to solving a nonlinear integer programming problem. We prove the continuous relaxation version is convex, and use the Karush-Kuhn-Tucker (KKT) condition to find its global optimal solution. Finally the branch-and-bound method is utilized to find the binary solution of original problem.
Due to the page limit, we omit the details of the solution and refer the reader to \cite{li2022resource} for a similar example.

\begin{figure}[htb!]
\begin{minipage}{0.48\linewidth}
  \centerline{\includegraphics[width=4.9cm]{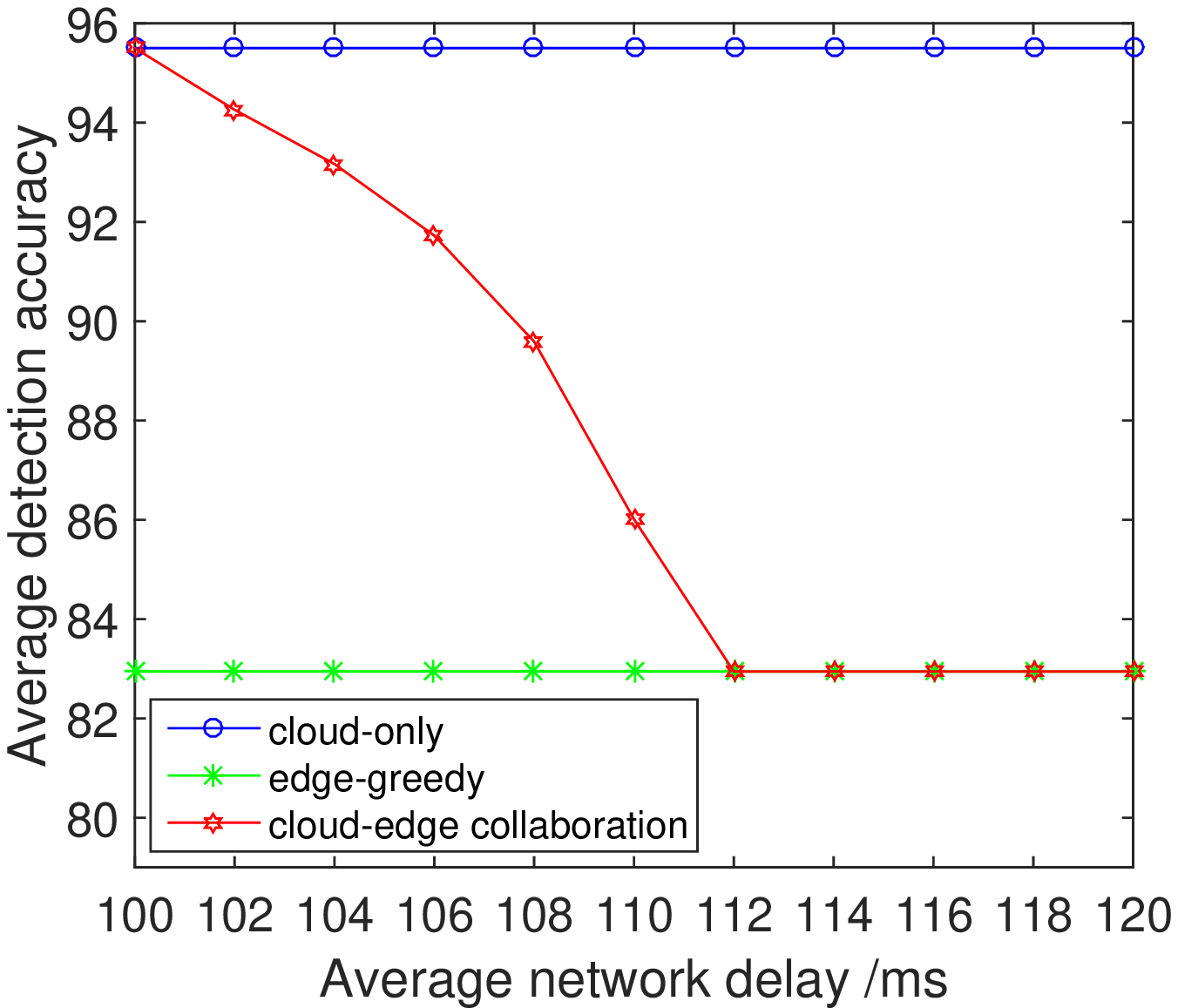}}
\centering
\center{(a)}
\end{minipage}
\hfill
\begin{minipage}{0.48\linewidth}
  \centerline{\includegraphics[width=4.9cm]{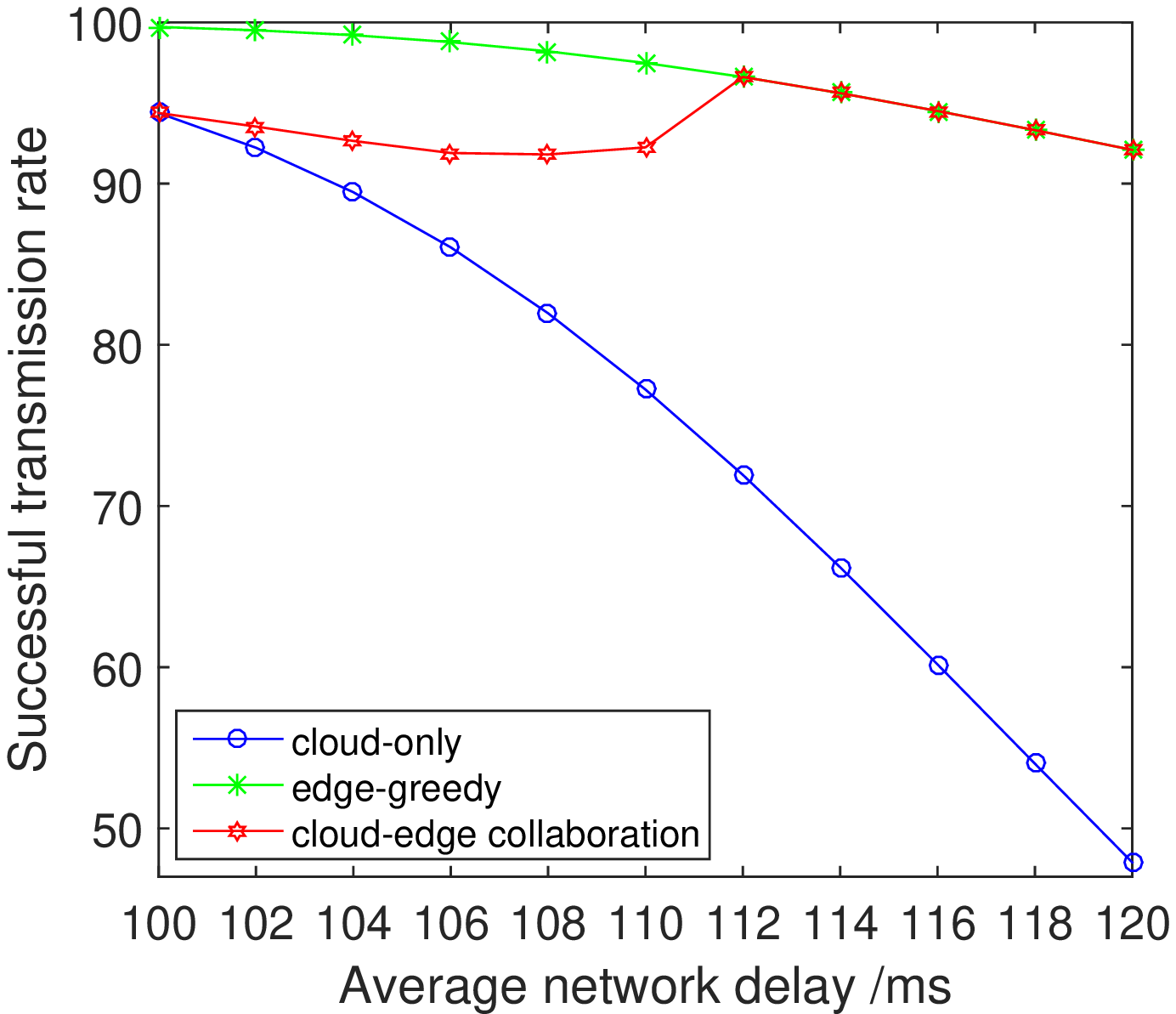}}
\center{(b)}
\end{minipage}
\vfill
\begin{minipage}{0.48\linewidth}
  \centerline{\includegraphics[width=4.9cm]{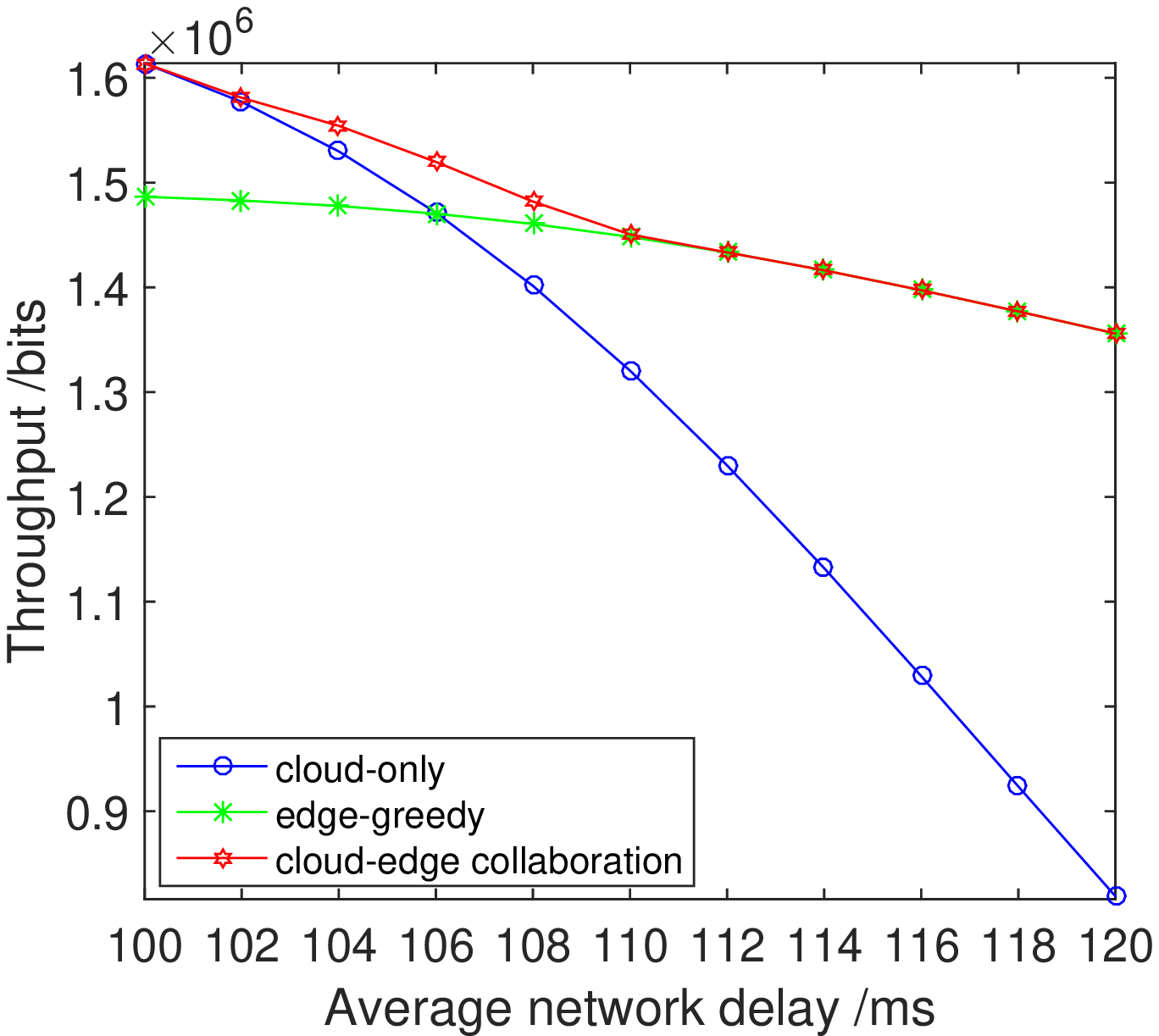}}
  \center{(c)}
\end{minipage}
\hfill
\begin{minipage}{0.48\linewidth}
  \centerline{\includegraphics[width=4.9cm]{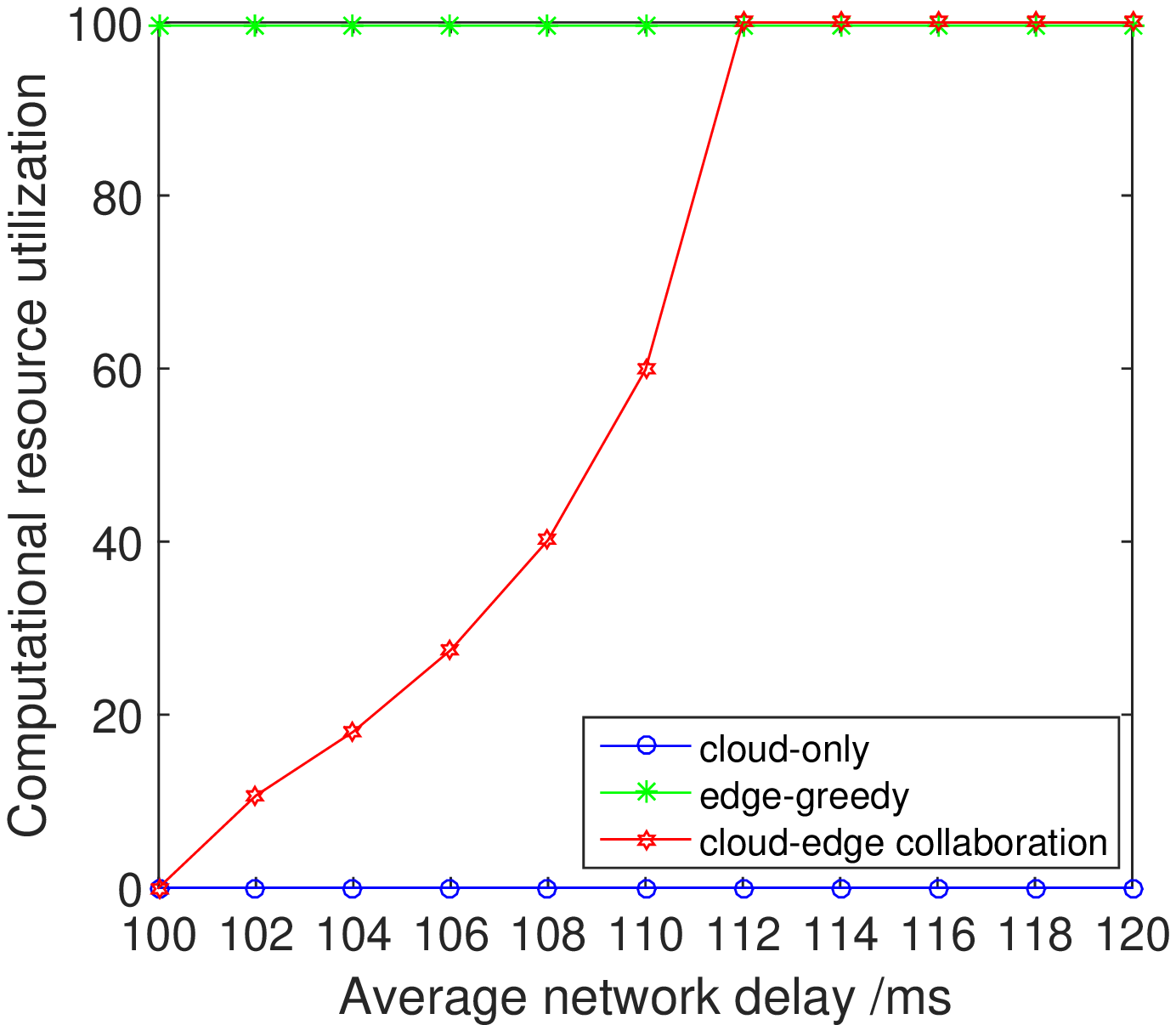}}
\center{(d)}
\end{minipage}
\caption{Performance under different average network delays. (a) Average detection accuracy. (b) Successful transmission rate. (c)  Throughput. (d) Computational resource utilization.}
\label{fig_changes_mu}
\end{figure}

\subsection{Experimental Results}
We have conducted experiments to compare the proposed scheme with the cloud-only (all data is transmitted to the cloud server for processing) scheme and the edge-greedy scheme (greedy scheduling of as much data as possible to the edge side, and the remaining data will be transmitted to the cloud platform), the result is shown in Fig. \ref{fig_changes_mu}. It can be seen that when the network delay is small, data trends to be transmitted to the cloud server for processing. The cloud server has greater computing power and can provide higher detection accuracy under the delay requirement. With the increase of network delay, data tends to be transmitted to edge devices. The transmission delay of the edge device is short and the successful transmission rate is high, but the detection accuracy is lower than that of the cloud server. Under the cloud-only and edge-greedy scheme, the amount of data transmitted to the cloud and edge is unchanged, so the average detection accuracy remains unchanged and the successful transmission rate decreases. The average detection accuracy of the cloud-edge collaboration scheme also gradually decreases. When the network delay is small, the data tends to be transmitted to the cloud, and when the network delay is large, the data tends to be transmitted to the edge device, so the successful transmission rate of the cloud-edge collaboration scheme first decreases and then increases. Until the computational resources of the edge device are exhausted, excess data can only be transmitted to the cloud for processing, so the accuracy of the cloud-edge collaboration scheme remains unchanged, and the successful transmission rate decreases. With the increase of network delay, the throughput of cloud-only, edge-greedy and cloud-edge collaboration schemes all gradually decrease, but cloud-edge collaboration systems have the largest throughput.
%我们进行了仿真实验，将提出的云边协作方案与仅云（所有数据传输到云服务器进行处理）方案和边缘贪婪方案（贪婪调度尽可能多的数据到边缘侧，其余数据将传输到云平台），结果如图 \ref{fig_changes_mu} 所示。从图中可以看出，当网络延迟较小时，数据趋向于传输到云服务器进行处理。云服务器具有更大的计算能力，可以在延迟要求下提供更高的检测精度。随着网络延迟的增加，数据趋向于向边缘设备传输。边缘设备传输时延短，传输成功率高，但检测精度低于云服务器。在cloud-only和edge-greedy方案下，传输到云端和边缘的数据量不变，因此平均检测精度保持不变，传输成功率降低。云边协同方案的平均检测精度也逐渐降低。当网络延迟较小时，数据倾向于传输到云端，当网络延迟较大时，数据倾向于传输到边缘设备,因此，云边协同方案的传输成功率先下降后上升。直到边缘设备的计算资源耗尽，多余的数据只能传输到云端进行处理，所以云边协同方案的准确性保持不变，传输成功率下降。随着网络延迟的增加，纯云、边缘贪婪和云边协同方案的吞吐量均逐渐下降，但云边协同系统的吞吐量最大。

\section{Future work}
\label{future_work}

We envision that the autonomous smart grid fault detection will play a vital role in future society by offering a green, safe and efficient smart grid.
This area opens up many
exciting and critical future research directions.

\subsection{Standardization}
Smart grid requires a large number of on-site measurement and control devices of different models and functions to be widely deployed in the whole network. If there is no unified requirement in design, production and use, different measurement and control devices may adopt different information formats and communication protocols. The collected information can only meet the application requirements of local sub-nets or some functional systems, but cannot be applied by all functional systems in the entire smart grid. This is exactly a key factor affecting autonomous smart grid fault detection. Therefore, it is urgent to carry out standardization work.

The success of autonomous smart grid fault detection cannot be achieved without standardization, through the implementation and development of technical standards based on consensus of all parties, including companies, users, interest groups, standards organizations, and governments. %\textcolor{red}{liu:这部分如果你们能扩展下就更好了}
To the authors' knowledge, this part of the work is still in progress and the related standardization efforts need further work.

\subsection{Network for Artificial Intelligence}
%\textcolor{red}{liu:如果篇幅不够，这个network for AI 感觉可以删除}
Artificial intelligence algorithms can be used to deeply integrate and analyze the massive online and offline multi-source heterogeneous data of power systems, and extract the mapping relationships, key information and judgment rules to establish data-driven prediction, evaluation and diagnosis models, thus improving the response time and accuracy of system fault diagnosis and prediction.

To better support the application of artificial intelligence in autonomous smart grid fault detection, smart networks are expected.
There are several studies on improving  networks for artificial intelligence, but there are still many unresolved issues in autonomous smart grid fault detection, from the aspects of network architecture, network protocol to system optimization objective.

\subsection{Security}
By offloading the sensor data to edge devices or cloud servers, smart grid poses the problems of leaking node information, which may cause security problems. Security concerns for smart grid are discussed in  recent study \cite{8373734}. Particular issues of security assurance and privacy preservation in autonomous smart grid fault detection
must be further investigated as well.

In addition, distributed deep learning has recently become a research hotspot that allows learning without sharing raw data and provides an option for secure automated smart grid fault detection.

\subsection{Integration with Emerging Techniques}

UAV systems are generating increasing interest in academia and industry.
Due to their mobility, flexibility and maneuverability, they are effective options for a variety of uses such as surveillance and mobile edge computing \cite{liu2021robust}.
Network function virtualization (NFV) and software defined networking (SDN) enable the creation of automated, scalable and customizable networks.
Caching facilitates popular content delivery and dramatically reduces network load. In-network computing enables the execution of programs in network devices that normally run on end hosts. That is, in-network computing focuses on computation within the network, using devices that already exist in the network system and have been used to forward traffic.
These emerging technologies are already being used in smart grid systems and have improved their performance.

However, to the best of our knowledge, there are still many unresolved issues for optimal integration of these technologies in autonomous smart grid fault detection, such as the joint investigation of caching, task offloading with non-negligible input/output sizes, and aerial computing constraints.

%\textcolor{red}{
%\subsection{Real-time fault detection}
%The basic sensing elements of future smart grid are smart meters and advanced metering infrastructures. Relying mainly on the communication network, fault management is essential to reduce the synchronization problem and to improve the system stability. Nevertheless, most fault management techniques demand a data processing center to monitor and analyze the operation of the system in real-time, as well as fault indicators, local automation, and communication devices. Additionally, it is important to detect and locate the faults rapidly enough to avoid a complete collapse of smart grid system. The focus of future work is to achieve fast and accurate fault detection.
%}

\section{Conclusion}
\label{sec_conclusion}
This paper focuses on the autonomous fault detection in smart grids. We first present the basic principles of smart grid fault detection. Then we explain the new requirements for autonomous smart grid fault detection, the technical challenges and their possible solutions. A case study is conducted and shows the great advantages of the proposed solution. Furthermore, we highlight relevant directions for future research.

\section*{Acknowledgment}
This work is supported in part by grants from the National Natural Science Foundation of China (52077049, 51877060, 62173120), the Anhui Provincial Natural Science Foundation (2008085UD04), the 111 Project (BP0719039).

\bibliographystyle{IEEEtran}
\bibliography{reference}

\end{document}